\documentclass[a4paper,11pt]{article}

     \PassOptionsToPackage{numbers, compress}{natbib}
\usepackage[preprint]{neurips_2019}

\usepackage[utf8]{inputenc} % allow utf-8 input
\usepackage[T1]{fontenc}    % use 8-bit T1 fonts
\usepackage{hyperref}       % hyperlinks
\usepackage{url}            % simple URL typesetting
\usepackage{booktabs}       % professional-quality tables
\usepackage{amsfonts}       % blackboard math symbols
\usepackage{nicefrac}       % compact symbols for 1/2, etc.
\usepackage{microtype}      % microtypography

\usepackage{epsfig}
\usepackage{graphicx}
\usepackage{amsmath}
\usepackage{amssymb}
\usepackage{subfigure}
\usepackage{multirow}
\usepackage{array}
\usepackage{algorithm}
\usepackage{algorithmic}
\usepackage[utf8]{inputenc}

\graphicspath{{./figures/}}
\newcommand{\y}{{\boldsymbol y}}
\newcommand{\x}{{\boldsymbol x}}

\newcommand{\n}{{\boldsymbol n}}

\newcommand{\setF}{{\mathcal F}}
\newcommand{\setA}{{\mathcal A}}
\newcommand{\setP}{{\mathcal P}}

\title{Subsampled Fourier Ptychography using Pretrained Invertible and Untrained Network Priors}

\author{%
  Fahad Shamshad, Asif Hanif, and Ali Ahmed\\%\thanks{Use footnote for providing further information
  Department of Electrical Engineering,
  Information Technology University, Lahore, Pakistan\\
  \texttt{$\lbrace$fahad.shamshad, msee18003, ali.ahmed$\rbrace$@itu.edu.pk} \\
}

\begin{document}

\maketitle
\vspace{-1em}
\begin{abstract}
Recently pretrained generative models have shown promising results for subsampled Fourier Ptychography (FP) in terms of quality of reconstruction for extremely low sampling rate and high noise. However, one of the significant drawbacks of these pretrained generative priors is their limited representation capabilities. Moreover, training these generative models requires access to a large number of fully-observed clean samples of a particular class of images like faces or digits that is prohibitive to obtain in the context of FP. In this paper, we propose to leverage the power of pretrained invertible and untrained generative models to mitigate the representation error issue and requirement of a large number of example images (for training generative models) respectively. Through extensive experiments, we demonstrate the effectiveness of proposed approaches in the context of FP for low sampling rates and high noise levels.

%However, the representation capabilities of these pretrained generators do not capture the full distribution for complex classes of images, such as human faces or numbers, resulting in representation error.
 %Moreover, recent studies have shown that these pretrained generative priors struggle at high-resolution in imaging inverse problems for reconstructing a faithful estimate of the true image, potentially due to mode collapse issue. To mitigate the issue of representation error of pretrained generative models for subsampled FP, we propose to make pretrained generator image adaptive by modifying it to better represent a single image (at test time) that is consistent with the subsampled FP measurements. 

%This paper proposes a novel framework to regularize the highly ill-posed and non-linear Fourier ptychography problem using generative models. We demonstrate experimentally  that our proposed  algorithm, \textit{Deep Ptych}, outperforms the existing Fourier ptychography techniques, in terms of quality of reconstruction and robustness against noise, using far fewer samples. We further modify the proposed approach to allow the generative model to explore solutions outside the range, leading to improved performance.
\end{abstract}
\section{Introduction}
Resolution loss in long distance imaging can primarily be attributed to the diffraction blur, that is caused by limited aperture of the imaging system \cite{holloway2017savi}. To mitigate the effects of the diffraction blur, recently an emerging computational imaging technique known as Fourier Ptychography (FP) has shown promising results \cite{holloway2016toward,zheng2013wide}. The objective of FP is to recover a high-resolution image from multiple diffraction-limited low-resolution images. In this paper, we consider recovering the signal $\x \in \mathbb{R}^n$ captured via forward acquisition model of FP, given by:
\begin{equation} \label{eq:pr}
\y_\ell = \vert \setA_\ell (\x) \vert + \n_\ell, \;\; \text{for} \;\; \ell = 1,2,...,L,
\end{equation}
where $\y_\ell \in \mathbb{R}^m$ is diffraction-limited image corresponding to $\ell^{th}$ camera, $\setA_\ell:\mathbb{R}^n \rightarrow \mathbb{C}^m$ is the linear operator representing the forward acquisition model, and $\n_\ell \in \mathbb{R}^m$ denotes  noise perturbation. For $\ell^{th}$ camera, the linear operator $ \setA_\ell$ has the form  $ \setF^{-1} \setP_\ell {\circ} \setF$, where $\setF$ denotes 2D Fourier transform, $\setP_\ell$ is a pupil mask that acts as a bandpass filter in the Fourier domain, and $\circ$ represents the Hadamard product. Specifically, FP works by iteratively stitching together a sequence of frequency limited low-resolution images $\y_\ell$ in Fourier domain to recover the high-resolution true image $\x$. Since optical sensors can measure only the magnitude of the signal \cite{shechtman2015phase}, phase information is lost during the acquisition process --- making the FP problem highly ill-posed.

To make the FP problem well-posed, generally 
%generally additional information in form of overcomplete set of measurements are required i.e. $m \gg n$. These 
additional measurements (i.e. $m \gg n$) are acquired in the form of high overlapping frequency bands in the frequency domain \cite{zheng2013wide}. Although effective, these redundant measurements can pose severe limitations in terms of %data storage, specially at higher resolutions, and results in 
high computational cost.
Recently, by devising realistic sampling strategies, prior information (sparsity and structured sparsity \cite{jagatap2018sub}) about the true signal has been leveraged to reduce the number of measurements (subsampling) in FP setup. However, it has been observed that these conventional signal priors often fail to capture the rich structure that many natural signals exhibit \cite{hand2017global}. Priors learned from huge datasets can effectively capture this rich structure using the power of deep neural networks \cite{lecun2015deep}. These deep neural networks or deep learning based end-to-end approaches have not yet been explored for reducing the number of measurements in FP. Moreover, even a slight change
in the parameters of FP forward acquisition model such as number of cameras or overlap would require costly retraining of these models.

To bridge the gap between deep learning based approaches (that can take advantage of the powerful learned priors) and conventional \textit{hand-designed} priors such as sparsity (that are flexible enough to handle a variety of model parameters), recently pretrained deep generative models have emerged as an impressive alternative for solving inverse imaging problems including subsampled FP problem \cite{shamshad2019deep, shamshad2018robust, hand2018phase, shamshad2019adaptive}. However, one of the significant drawbacks of these pretrained generative priors is their limited representation capabilities. That is once trained, these pretrained generative models are incapable of producing any target image that lies outside their range\footnote{Range of the pretrained generator can be defined as the set of all the images that can be generated by the that generator.}. Moreover, training of these generative models require access to a large number of fully-observed clean
samples of a particular class of images like faces or digits. Unfortunately, obtaining multiple high-resolution samples can be expensive or impractical for many applications including FP.

In this paper, we aim to handle the aforementioned issues in context of subsampled FP by leveraging upon two recent works related to solving inverse imaging problems via pretrained invertible \cite{asim2019invertible} and untrained generative model \cite{ulyanov2018deep,heckel2018deep, heckel2019regularizing}. Our first contribution is to leverage the power of pretrained invertible generative models to mitigate the representation error issue of conventional generative models for non-convex and non-linear inverse problem of subsampled FP. We refer our first approach as \textit{Invertible Ptych}. Our second contribution is to relax the requirement of massive amount of training data for training of generative models which can be prohibitively expensive to obtain in domains such as FP. We refer to our second approach throughout this paper as \textit{Untrained Ptych}. Through numerical simulations, we demonstrate that proposed approaches can get better reconstructions, both qualitatively and quantitatively, at low subsampling ratios and high noise perturbations.

\section{Problem Formulation}
We refer interesting readers to related work of \cite{shamshad2018deep} for details of forward acquisition model of subsampled FP. As shown in \cite{jagatap2018sub, shamshad2018deep}, in order to reduce the sample complexity, we can discard some observations in the sensor plane. This can be treated as applying a subsampling operator ($\mathcal{M}_{\ell}(\cdot)$), to effectively reduce the number of measurements. Mathematically, observation $\boldsymbol{y}_\ell$ for $\ell$th camera can be modeled as%are given by {\color{blue}the} model {\color{blue}as shown below}
\begin{equation} \label{eq:forward-model}
\boldsymbol{y}_\ell = \vert \mathcal{M}_\ell (\mathcal{A}_\ell (\boldsymbol{x}))\vert + \boldsymbol{n}_\ell,
\end{equation}
where $\mathcal{A}_\ell = \mathcal{F}^{-1} \mathcal{P}_\ell \circ \mathcal{F}$ is the measurement model prior to optical sensor acquisition step, $\circ$ denotes the Hadamard product, and $\mathcal{M}_\ell$ is the subsampling operator. Subsampling operator when applied to measurements $\boldsymbol{y}$, randomly picks a fraction of samples $(f)$ discarding the others \cite{jagatap2018sub}. We define the subsampling ratio as the fraction of samples retained by $\mathcal{M}_\ell$ divided by the total number of observed samples i.e. 
\begin{equation}
\footnotesize
\text{Subsampling Ratio (\%)} =\frac{\text{Fraction of samples retained ($f$)}\times 100}{\text{Total observed samples ($nL$)}}. \nonumber 
\end{equation}
The subsampling mask resembles the operation of a binary  matrix having entries 1's and 0's. The mask has been element-wise multiplied with the observations in such a way that pixels corresponding to 1's are retained and those corresponding to 0's are discarded. Hence subsampling ratio governs the percentage of samples that will be retained. Without assuming any prior information about the true image $\boldsymbol{x}$, we can minimize the FP measurement loss as 
\begin{equation} \label{eq:reconstruction_constraint}
\hat{\boldsymbol{x}} = \underset{\boldsymbol{x} \in \mathbb{R}^n}{\text{arg min}} \sum_{\ell=1}^{L} \Vert \boldsymbol{y}_\ell - \vert \mathcal{M}_\ell\mathcal{A}_\ell(\boldsymbol{x}) \vert \Vert^{2}_2,
\end{equation}
to find the estimate of the true image $\boldsymbol{x}$. Note that without assuming any prior information about $\boldsymbol{x}$, \eqref{eq:reconstruction_constraint} is notoriously difficult to solve as infinitely many solutions satisfy \eqref{eq:reconstruction_constraint}.
\vspace{-1em}
\section{Proposed Approach}
In this section, we formally introduce our proposed approaches \textit{Untrained Ptych} and \textit{Invertible Ptych}. Specifically, we denote by $G_\theta(z)$ as a deterministic function representing generative model that takes input $z \in \mathbb{R}^k$ and parameterized by the set of weights $\theta \in \mathbb{R}^d$ to produce output $G_\theta(z) \in \mathbb{R}^n$.

\textbf{Untrained Ptych}: Following the work of \cite{ulyanov2018deep}, we propose to use the structure of untrained convolutional generative models as a prior to obtain estimate of true image $\boldsymbol{x}$. Specifically, \textit{Untrained Ptych} approach aims to find the set of weights of convolutional generative model (initialized randomly), $\hat{\theta}$, that produce output $G_{\hat{\theta}}(z)$ which best matches with the subsampled FP measurements $\boldsymbol{y}$ while obeying forward acquisition model  \eqref{eq:reconstruction_constraint}. Specifically, we solve the following optimization problem
\begin{equation} \label{eq:laten_optimization}
\hat{\theta} = \underset{\theta \in \mathbb{R}^d}{\text{arg min}} \sum_{\ell=1}^{L} \Vert \boldsymbol{y}_\ell - \vert \mathcal{M}_\ell\mathcal{A}_\ell({G}_{\theta}(\boldsymbol{z})) \vert \Vert^{2}_2.
\end{equation}
%such that $\hat{\boldsymbol{x}} = \mathcal{G}_{\theta}({\hat{\boldsymbol{z}})}$ is the reconstructed signal. This optimization program modifies the latent representation vector $\boldsymbol{z}$ such that the generator generates an image $G_{\theta}(\hat{\boldsymbol{z}})$ that is consistent with (??). 
The estimated image $\hat{\boldsymbol{x}}$ is acquired by a forward pass of the $z$ through the
generator ${G}_{\hat\theta}$ as $\hat{\boldsymbol{x}} = {G}_{\hat{\theta}}({z})$.

\textbf{Invertible Ptych}: For \textit{Invertible Ptych}, we assume that we have access to a pretrained invertible generative model $G_\theta()$ ($\theta$ denotes pretrained weights) that has been trained on a specific class of natural images like face dataset. In this work, we use invertible architecture GLOW \cite{kingma2018glow}. For \textit{Invertible Ptych}, we solve the following optimization program via gradient descent
\begin{equation} \label{eq:laten_optimization}
\hat{z} = \underset{z \in \mathbb{R}^k}{\text{arg min}} \sum_{\ell=1}^{L} \Vert \boldsymbol{y}_\ell - \vert \mathcal{M}_\ell\mathcal{A}_\ell({G}_{\theta}(\boldsymbol{z})) \vert \Vert^{2}_2.
\end{equation} 
The estimated image $\hat{\boldsymbol{x}}$ is acquired by a forward pass of the $\hat{z}$ through the pretrained
generator ${G}_{\theta}$ as $\hat{\boldsymbol{x}} = {G}_{{\theta}}(\hat{z})$.

\vspace{-1em}
\section{Numerical Simulations}
\vspace{-1em}
\begin{figure}[t] 
\centering
\subfigure{\includegraphics[width=0.20\columnwidth]{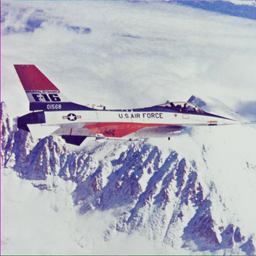}} %\hspace{-0.5em}
\subfigure{\includegraphics[width=0.20\columnwidth]{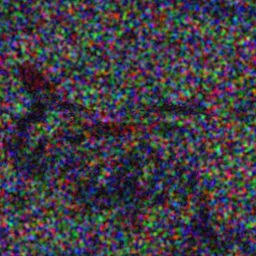}} %\hspace{-0.5em}
\subfigure{\includegraphics[width=0.20\columnwidth]{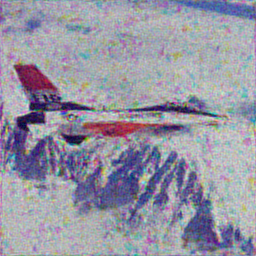}} %\hspace{-0.5em}
\subfigure{\includegraphics[width=0.20\columnwidth]{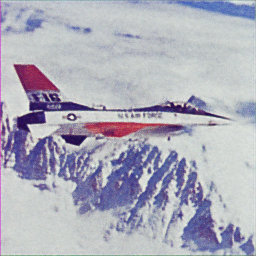}} \\[-0.8em]
\setcounter{subfigure}{0}
\subfigure[Original]{\includegraphics[width=0.20\columnwidth]{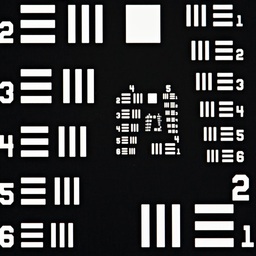}}\hspace{0.05em}
\subfigure[IERA]{\includegraphics[width=0.20\columnwidth]{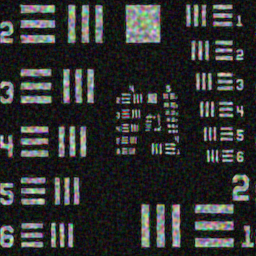}} %\hspace{-0.5em}
\subfigure[CoPRAM]{\includegraphics[width=0.20\columnwidth]{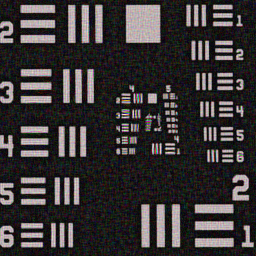}} %\hspace{-0.5em}
\subfigure[Untrained Ptych]{\includegraphics[width=0.20\columnwidth]{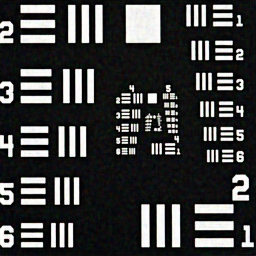}} \\[-0.8em]
\setcounter{subfigure}{0}

%\subfigure{\includegraphics[width=0.179\columnwidth]{./figures/invertible/subsamp/original_images/009.png}} \hspace{-0.5em}
%\subfigure{\includegraphics[width=0.179\columnwidth]{./figures/invertible/subsamp/iera/2/009.png}} \hspace{-0.5em}
%\subfigure{\includegraphics[width=0.179\columnwidth]{./figures/invertible/subsamp/copram/2/009.png}} \hspace{-0.5em}
%\subfigure{\includegraphics[width=0.179\columnwidth]{./figures/invertible/subsamp/dcgan/2/009.png}}
%\subfigure{\includegraphics[width=0.179\columnwidth]{./figures/invertible/subsamp/invertible/2/009.png}} \\[-0.8em]
\subfigure[Original]{\includegraphics[width=0.167\columnwidth]{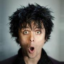}} \hspace{-0.5em}
\subfigure[IERA]{\includegraphics[width=0.167\columnwidth]{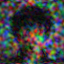}} \hspace{-0.5em}
\subfigure[CoPRAM]{\includegraphics[width=0.167\columnwidth]{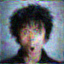}} \hspace{-0.5em}
\subfigure[DCGAN]{\includegraphics[width=0.167\columnwidth]{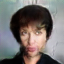}}
\subfigure[Invertible]{\includegraphics[width=0.167\columnwidth]{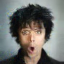}} \\[-0.8em]

\caption{\small First two rows show subsampled FP reconstruction results of  \textit{Untrained Ptych} along with baseline methods. First row shows results with 2$\%$ subsampling ratio in noiseless setting. Second row shows result with 5$\%$ noise for subsampling ratio of 20$\%$. Last row shows results of \textit{Invertible Ptych} reconstructions for sampling rate of 2$\%$ in noiseless setting. Reconstructions of proposed approaches are visually appealing and relatively less corrupted with artifacts as compared to that of CoPRAM and IERA. }
\label{fig:untrained_subsamp}
\end{figure}
In this section, we evaluate the performance of proposed approaches on trained and untrained generative networks for different subsampling ratios and noise levels\footnote{Noise of 1$\%$, for image scaled between 0 to 1, translates to Gaussian noise with zero mean and a standard deviation of 0.01}. To quantitatively evaluate the performance of our algorithm, we use two metrics Peak Signal to Noise Ratio (PSNR) and Structural Similarity Index Measure (SSIM). %In all our experiments, we report results on a held out test set, unseen by the generative models during training. Our proposed approaches has been implemented on the Tensorflow framework on an Nvidia Titan X GPU. 

\textbf{Comparison Methods} We consider IERA (Iterative Error Reduction Algorithm) \cite{holloway2016toward}, CoPRAM \cite{jagatap2018sub} and Deep Ptych \cite{shamshad2018deep} as our baseline methods for evaluating the performance of our approaches for different subsampling ratios and noise levels.% IERA solves FP problem by alternatively enforcing spatial and Fourier domain constraints whereas CoPRAM exploits the underlying sparse structure of true signal for faithful reconstruction at low subsampling ratios. For CoPRAM, we assume sparsity of images in the Fourier domain. We use the default algorithmic parameters for all comparison methods, unless stated otherwise.

\textbf{Architectures} For untrained generative model, we use U-Net based architecture with skip connections as in \cite{ulyanov2018deep}. For invertible generative model, we use the Glow architecture \cite{kingma2018glow}, though our framework could be used with other invertible architectures as well. Let $K$ be the number of steps of flow before a splitting layer, and $L$ be the number of times the splitting is performed. To train over CelebA\cite{lecun2015deep}, we choose the network to have $K= 32$, $L= 4$ and affine coupling, and train it with a learning rate $0.0001$.%We specifically model our invertible networks after the recently proposed Glow architecture, which consists of a multiple flow steps. Each flow step comprises of an activation normalization layer, a $1 \times 1$ convolutional layer, and an affine coupling layer, each of which is invertible. Let K be the number of steps of flow before a splitting layer, and L be the number of times the splitting is performed. To train over CelebA, we choose the network to have K= 32, L= 4 and affine coupling, and train it with a learning rate $0.0001$, and a batch size of 12. 
The model was trained over 8-bit images with 10,000 warmup iterations as in \cite{kingma2018glow}. We refer the reader to \cite{kingma2018glow} for specific details on the operations performed in each of the network layer.

%Deep convolutional generative adversarial network (DCGAN) is trained on this rescaled dataset with architecture as proposed in \cite{radford2015unsupervised}  having latent dimension size set to 100, batch size of 64, $\beta_1$ set to 0.5, $\lambda$ set to  $1.4 \times 10^{-4}$, and learning rate of $2 \times 10^{-4}$ using Adam optimizer. DCGAN model is trained by updating generator $G$ twice and discriminator D once in each cycle to avoid fast convergence of $D$.

\textbf{Datasets} For experiments with natural images, we perform subsampling and noise experiments on 5 standard test images of Aeroplane, Cameraman, Lena, USAF chart, and Boat each with size of 256 $\times$ 256. For face images, Glow model has been trained on CelebA dataset, center cropped to size  64$\times$64 %as shown in Figure \ref{fig:test_images}. 
We use 30,000 images for training Glow and DCGAN(for Deep Ptych) based generative models \cite{radford2015unsupervised} . 
\vspace{-1.9em}
\subsection{Qualitative and Quantitative results}
In this section we provide the qualitative and quantitative results of subsampled FP for \textit{invertible-ptych} and \textit{untrained-ptych} by varying the subsampling ratio and noise level.

Qualitative results for \textit{untrained-ptych} for subsampling ratio of 1$\%$ (in noiseless setting) and noise level of 5$\%$ (for 20$\%$ subsampling ratio) are shown in Figure \ref{fig:untrained_subsamp}. It can be seen that reconstructed images via \textit{untrained ptych} are visually appealing as compared to those of IERA and CoPRAM that contain artifacts. We observe similar trend for \textit{invertible ptych} as shown in Figure \ref{fig:untrained_subsamp} (bottom row). Note that the results of DCGAN, though sharp, are constrained to lie in the range of the pretrained generator. On the other hand, \textit{invertibe ptych} has no such limitation (by design).

\begin{figure}[t]
	\vspace{-1.8em}
\centering
%\subfigure[\scriptsize ]{\includegraphics[trim={0.45cm 0.0cm 0.1cm 0.3cm},clip,width=.5\columnwidth]{./figures/ssim_subsamp_untrained.eps}} \hspace{-0.28cm}

\subfigure[\scriptsize Untrained Ptych]{\includegraphics[trim={0.45cm 0.1cm 0.1cm 0.3cm},clip,width=.43\columnwidth]{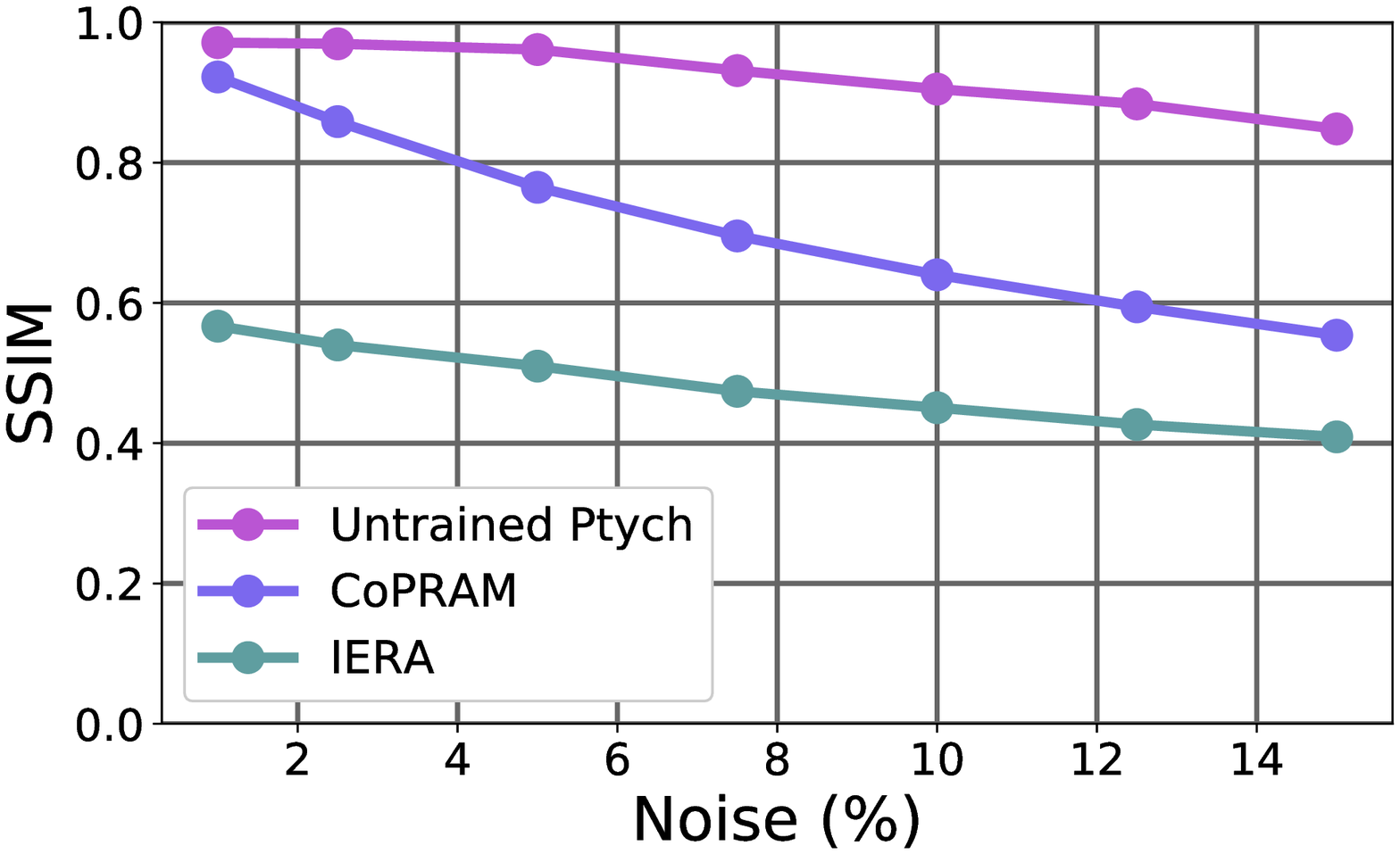}} \hspace{-0.15cm}
\subfigure[\scriptsize Invertibe Ptych ]{\includegraphics[trim={2.65cm 0.0cm 0.1cm 0.3cm},clip,width=.38\columnwidth]{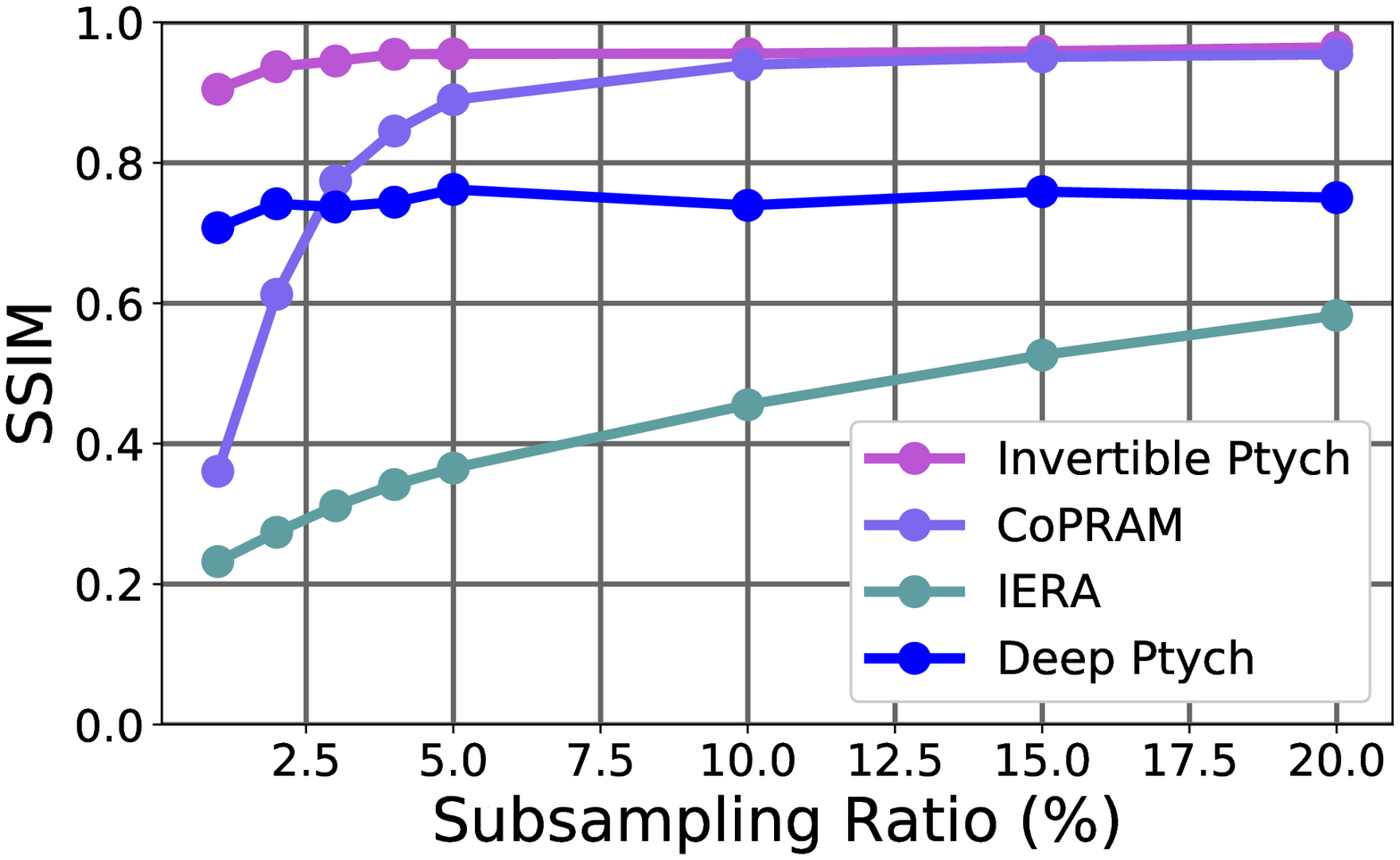}} 
\vspace{0.1em}
%\subfigure[\scriptsize ]{\includegraphics[trim={2.5cm 0.1cm 0.1cm 0.3cm},clip,width=.5\columnwidth]{./figures/ssim_noise_invertible.eps}} \hspace{-0.3cm}
\caption{ \small {Average SSIM plots of \textit{Untrained Ptych} and \textit{Invertible Ptych} for different noise levels and subsampling ratios.}}
 \label{fig:quatitative_plots}
\end{figure}

Quantitative results, in terms of PSNR and SSIM, for \textit{Untrained Ptych} and \textit{Invertible Ptych} are shown in Figure \ref{fig:quatitative_plots} and Table \ref{table:performance} respectively. The results are averaged over 5 images for \textit{Untrained Ptych} and 15 CelebA test images for \textit{Invertible Ptych}. As shown in Figure \ref{fig:quatitative_plots}, \textit{Untrained Ptych} is able to acheive higher SSIM values for all noise levels as compared to IERA and CoPRAM. Similar performance gain has been observed for \textit{Invertibe Ptych} for roustness against noise as compared to baseline methods. PSNR results in Table \ref{table:performance} indicate that proposed approaches are able to acheive considerable gain for low subsampling ratios and high noise levels as compared to baseline methods.

\begin{table}[t]
	\caption{\small{PSNR (dB) for different subsampling ratios and noise levels for \textit{Untrained Ptych} and \textit{Invertible Ptych}}}
	\vspace{-1em}
	\label{table:performance}

	\begin{center}
		\vspace{0.2em}

		\resizebox{0.85\textwidth}{!}{
			\begin{tabular}{c|ccccc|ccccc}

				\hline \hline
				
				\multicolumn{8}{c}{\textbf{Subsamp Ratio (\%)}} &  \multicolumn{1}{c}{\textbf{Noise (\%)}}\\

				\textbf{} & \textbf{1} & \textbf{2}  &  \textbf{3}  & \textbf{5} & \textbf{10} & \textbf{1}  &  \textbf{2.5} & \textbf{5} & \textbf{7.5} & \textbf{10}   \\
				
				\hline \hline
				
				\textbf{IERA}  & 7.92  & 8.83 & 9.3 & 10.47 & 12.45  &  14.43 & 14.62 & 14.66 & 14.39 & 13.96 \\
				\textbf{CoPRAM}  & 12.07  & 18.08 & 21.61 & 22.94 & 24.94  & 24.57 & 22.7 & 22.42 & 21.39 & 20.32 \\
				\textbf{Untrained}  & \textbf{23.53} & \textbf{25.53} & \textbf{28.46} & \textbf{30.01} & \textbf{30.67}   & \textbf{33.99}  & \textbf{33.83} & \textbf{33.01} & \textbf{30.2} & \textbf{29.51} \\
				\hline \hline
				\textbf{IERA} & 9.00 & 9.89 & 10.68 & 11.51 & 13.37 & 15.51 & 15.16 & 15.38 & 15.17 & 14.80 \\
                \textbf{CoPRAM} & 12.07 & 16.33 & 19.57 & 21.94 & 23.29 & 24.27 & 23.61 & 22.55 & 21.43 & 20.87 \\
                \textbf{Deep Ptych} & 19.92 & 20.61 & 20.29 & 20.91 & 20.44 & 20.72 & 20.87 & 20.87 & 20.99 & 20.79 \\
                \textbf{Invertible} & \textbf{25.41} & \textbf{27.42} & \textbf{28.36} & \textbf{29.23} & \textbf{29.71} & \textbf{30.60} &\textbf{ 28.64} & \textbf{27.73} & \textbf{28.01} & \textbf{27.29} \\
	
				\hline \hline

			\end{tabular} 
		}
	\end{center}
	\vspace{-1em}
\end{table}

\section*{Acknowledgement} \nonumber

We gratefully acknowledge the support of the NVIDIA Corporation for the donation of NVIDIA TITAN Xp GPU for this research.

%\begin{figure}[t]
%\centering
%{\includegraphics[trim={0.45cm 0.0cm 0.1cm 0.3cm},clip,width=.8\columnwidth]{./figures/Unet.jpg}} 
% \label{fig:quatitative_plots}
%\end{figure}
\bibliographystyle{unsrt}
\bibliography{refs}
\end{document}